# Spatial Distribution of Element Abundances and Ionization States in Solar Energetic-Particle Events

**Donald V. Reames**

Institute for Physical Science and Technology, University of Maryland, College Park, MD 20742-2431 USA, email: dvreames@umd.edu

**Abstract** We have studied the spatial and temporal distribution of abundances of chemical elements in large "gradual" solar energetic-particle (SEP) events, and especially the source plasma temperatures, derived from those abundances, using measurements from the *Wind* and *Solar TErrestrial RElations Observatory* (STEREO) spacecraft, widely separated in solar longitude. A power-law relationship between abundance enhancements and mass-to-charge ratios [*A/Q*] of the ions can be used to determine *Q*-values and source plasma temperatures at remote spacecraft with instruments that were not designed for charge-state measurements. We search for possible source variations along the accelerating shock wave, finding one clear case where the accelerating shock wave appears to dispatch ions from 3.2±0.8 MK plasma toward one spacecraft and those from 1.6±0.2 MK plasma toward another, 116$^o$ away. The difference persists three days and then fades away. Three other SEP events show less-extreme variation in source temperatures at different spacecraft, in one case observed over 222$^o$ in longitude. This initial study shows how the power-law relation between abundance enhancements and ion *A/Q*-values provides a new technique to determine *Q* and plasma temperatures in the seed population of SEP ions over a broad region of space using remote spacecraft with instruments that were not originally designed for measurements of ionization states.

# 1. Introduction

The relative abundances of the chemical elements and isotopes have been a major factor in distinguishing energetic-particle sources and in understanding the physical processes of their acceleration and transport (*e.g.* Reames, 1999). For solar energetic particle (SEP) events, there has been a long history distinguishing the small "impulsive" events, with 1000-fold enhancements of $^3$He/$^4$He and of heavy elements ($Z > 50$)/O, associated with magnetic reconnection at the Sun in flares and jets, from the large "gradual" or long-duration events, where elements with abundances more like those of the solar corona are accelerated at shock waves driven out from the Sun by coronal mass ejections (CMEs) (see reviews by Gosling, 1993; Lee, 1997; Reames, 1999, 2013, 2015, 2017; Desai *et al.*, 2004; Mason, 2007; Kahler, 2007; Cliver and Ling, 2007, 2009; Kahler *et al.*, 2012; Wang *et al.*, 2012). The association of gradual events, the most intense and energetic of the SEP events, with CME-driven shocks has been well established (see Kahler *et al.*, 1984; Mason, Gloeckler, and Hovestadt, 1984; Kahler, 1992, 1994, 2001; Gopalswamy *et al.*, 2004, 2012; Cliver, Kahler, and Reames, 2004; Lee 2005; Rouillard *et al.*, 2011, 2012; Lee, Mewaldt, and Giacalone, 2012; Desai and Giacalone 2016), although, for a time, this association was complicated by the fact that these shocks can also reaccelerate residual suprathermal ions left over from previous impulsive SEP events (Mason, Mazur, and Dwyer 1999; Desai *et al.* 2003, 2006; Tylka *et al.*, 2005; Tylka and Lee, 2006).

## 1.1 Abundances and A/Q

Similarities of SEP abundances in large SEP events with those of the corona and solar wind were recognized early (*e.g.* Webber 1975; Meyer 1985) and Breneman and Stone (1985) found that the enhancement or depression of abundances in individual events varied as a power law in the charge-to-mass ratio [$Q/A$] of each element. As accelerated ions travel out along the magnetic field from the shock to an observer, they are scattered by magnetic irregularities, such as Alfvén waves. This scattering depends upon the magnetic rigidity, or momentum per unit charge, of the particle. If we compare different particle species at the same velocity, their difference in rigidity becomes a difference in $A/Q$. Thus an abundance ratio such as Fe/O will increase early in an event causing it to be de-pleted later since Fe, with a high ratio of *A/Q*, scatters less than O, with a lower ratio. Solar rotation turns this variation in time to an additional variation with solar longitude (see *e.g.* Reames 2013, 2015). In fact, streaming particles amplify Alfvén waves, greatly complicating the transport in the largest events with high proton intensities (see Ng, Reames, and Tylka 2003; Ng and Reames 2010). Fortunately, for most events, diffusion theory that assumes time-invariant scattering mean free paths [$\lambda$] which varies as a power of *A/Q*, seems adequate to explain (Reames 2016b) the temporal dependence of abundance ratios which vary as a power of *A/Q* that is linear in $t^{-1}$.

Thus if $\lambda_X$ of species X depends upon the particle magnetic rigidity [*P*] as a power law $P^\alpha$ and upon distance from the Sun [*R*] as $R^\beta$ we can express the ratio of the enhancement of X to that of O, as (Reames 2017, 2016a, b)

$$X/O \approx L^{\tau/t - 3/(2-\beta)} r^S \qquad (1)$$

where $L = \lambda_X / \lambda_O = r^\alpha = ((A_X/Q_X)/(A_O/Q_O))^\alpha$ and $\tau = 3R^2/[\lambda_O (2-\beta)^2 v]$ for particles of speed *v*. The factor $r^S$ represents any *A/Q*-dependent power-law enhancement at the source, prior to transport. For shock acceleration of impulsive suprathermal ions, it describes the power-law enhancement of the seed particles. For shock acceleration of the ambient coronal material, *S* = 0. In any case, log (X/O) varies linearly with log ($A_x/Q_x$).

The enhancement of abundances, relative to their coronal values depends upon *A/Q* of each species, but the values of *Q* depend upon the electron temperature of the source plasma, *T*. Can we find a value of *T* that determines *Q*-values that give the observed pattern of enhancements?

### 1.2 Abundances and Source Temperatures

An early attempt to deduce *T* from abundance enhancements was made by Reames, Meyer, and von Rosenvinge (1994). They noted that in impulsive SEP events C, N, and O were unenhanced, like He, and were probably also fully ionized with $A/Q \approx$ 2.0, while Ne, Mg, and Si formed a group that might correspond to a two-electron closed shell with $A/Q \approx 2.33$ and Fe had $A/Q \sim 3.6$. These ionization levels occur in a coronal temperature range of 3–5 MK. Using more accurate modern abundance measurements, which included the elements from $2 \leq Z \leq 82$, Reames, Cliver, and Kahler (2014a, 2014b, 2015) found that nearly all of 111 impulsive SEP events had source temperatures within 2.5–3.2 MK apparently from coronal active regions. Impulsive SEPs do *not* share flare temperatures, typically 10–20 MK where even Ne, Mg, and Si become fully stripped so that their abundances could not be enhanced relative to He, C, or O.

A graph of *A/Q vs. T* for a variety of elements is shown in Figure 1. As *T* changes, ionization states tend to linger at closed electron shells of an atom. As temperature decreases below 3 MK, first O, then N, then C, cross from the fully-ionized shell, like He to the two-electron "Ne-like" shell. In the same temperature excursion, Ar joins Ca, then S, Si, and then Mg cross to the "Ar-like" shell.

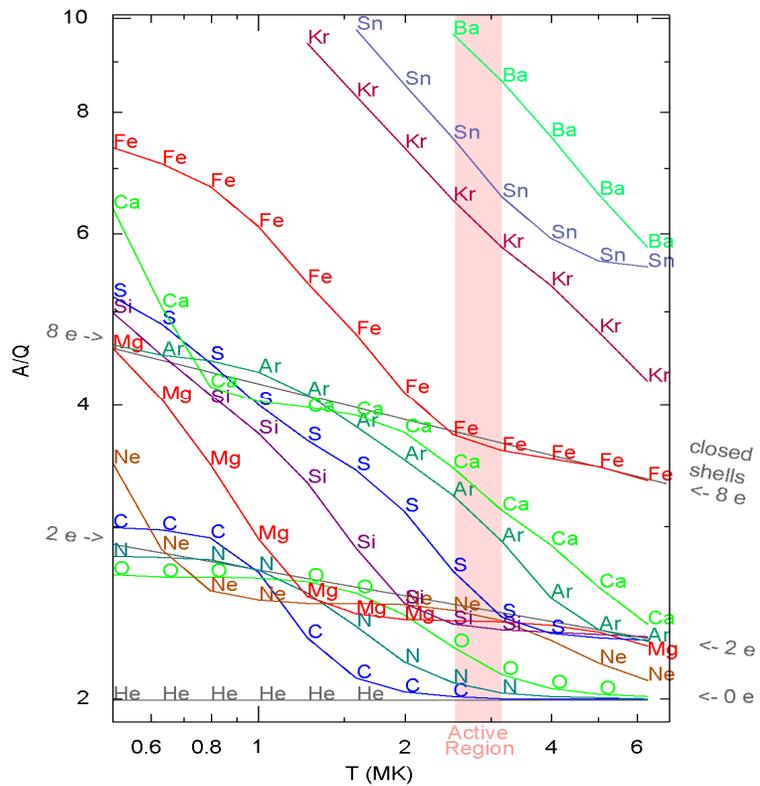

**Figure 1.** *A/Q* varies as a function of the theoretical equilibrium temperature for elements named along each curve. Data are from Mazzotta *et al.* (1998) up to Fe and from Post *et al.* (1977) above. Points are spaced every 0.1 of $\log_{10}(T)$. *A/Q*-values tend to cluster in bands produced by closed electron shells; those with 0, 2, and 8 electrons are shown, He having no electrons. Elements move systematically from one band to another as temperatures change. The shaded band corresponds to active-region temperatures found for impulsive SEP events.

If we wish to determine a temperature from observed enhancements, we remember that *A/Q* must have the same pattern as the enhancements and ask, is the abundance of C like He, like Ne, or between? Is Si like Ne, or like Ar? These patterns can uniquely determine the temperature. Procedurally we can least-squares fit log(X/O) *vs.* log(*A/Q*) at each temperature on the grid, record the value of $\chi^2$ at each *T*, and select the fit and the value of *T* with the lowest $\chi^2$. Examples of this type of analysis on gradual SEP events are shown and discussed in some detail by Reames (2016a) who found temperatures that were consistent for times throughout 45 gradual SEP events.

Using abundance enhancements and *A/Q* dependence to determine consistent source temperatures is a relatively new technique that can be applied widely in SEP events. For impulsive SEP events it provides temperatures at the point of acceleration, whereas direct charge measurements in space show modification of *Q* when ions pass through material when they leave sources below ~1.5 solar radii (*e.g.* DiFabio *et al.* 2008). For gradual events, mean values of $Q_{Fe}$ range from 10–14 as determined by direct and geomagnetic measurements (see review by Klecker 2013). These values are consistent with coronal temperatures of ~1.2–2.5 MK. However, measurements of SEP events are no longer available near Earth, much less at distant longitudes or for the rarer elements that we consider.

Instruments, at remote locations in the heliosphere, which can measure abundances of a large number of elements in SEP events, can now use this new technique to derive particle ionization states and source temperatures, a measurement that was never considered when these instruments were designed. Thus we can compare element abundances, probable ionization states, and source temperatures of SEP ions from the two spacecraft of the remote *Solar Terrestrial Relations Observatory* (STEREO) and the *Wind* spacecraft near Earth. Such widely distributed direct measurements of charge states and temperatures would not be possible otherwise.

### *1.3 Can SEP Abundances and Source Temperatures Vary with Longitude along a Shock?*

Particles from gradual SEP events often occupy a large fraction of the inner heliosphere. The spatial variation of proton and electron intensities and of their energy spectra have been studied widely within these events (*e.g.* Reames, Barbier, and Ng 1996; Reames, Ng, and Tylka 2012), especially with respect to the spectrally invariant region behind the CME called the reservoir (McKibben 1972; Roelof *et al.* 1992; Reames, Kahler, and Ng 1987; Diabog *et al.* 2003; Lario 2010). However, the spatial distribution of element abundances has primarily been studied by comparing event-averaged properties, such as Fe/O at different longitudes (*e.g.* Cohen *et al.* 2013).

However, we are left with more-detailed questions. Can the abundances or source temperatures of the seed population vary along the surface of the accelerating shock wave? Are such possible spatial differences erased by scattering or other modes of transport? Are variations erased across the entire spatial SEP distribution or primarily in the reservoir? We compare the full pattern of SEP abundances measured at widely separated platforms: STEREO-Ahead (A) and -Behind (B) Earth in solar orbit, and the *Wind* spacecraft near Earth. During several large SEP events, that intercept two or more spacecraft, we measure the relative abundances of the elements He, C, N, O, Ne, Mg, Si, S, Ar, Ca, and Fe using the *Low-Energy Matrix Telescope* (LEMT; von Rosenvinge *et al.* 1995) on *Wind* and the *Low-Energy Telescope* (LET; Mewaldt *et al.* 2008; see also Luhmann *et al.* 2008) on STEREO. Abundances are taken primarily from the 3.2–5 MeV amu$^{-1}$ interval on LEMT, and the 4–6 MeV amu$^{-1}$ interval of LET (S is not available on LET; data from http://www.srl.caltech.edu/STEREO/Level1/LET_public.html). Abundance enhancements are all measured relative to the average SEP abundances listed by Reames (2017; see also Reames 1995, 2014). With 10 or 11 elements to determine the two parameters (slope and intercept) of the linear fit of log (X/O) *vs.* log (*A/Q*) at each temperature on the grid (Figure 1), in most cases we have $m \approx 8$ degrees of freedom to follow the complex abundance patterns implied in Figure 1. However, since LET is much smaller than LEMT, it requires higher intensities for measurement of the rarer elements; this limits the selection of events.

### *1.4 Are SEP Sources Isothermal?*

As the solar wind flows out of the corona and the electron density decreases, the atomic electron capture and loss cross sections no longer allow changes in ionization states of atoms, causing "freezing in" of the ionization states. The freezing-in occurs as the ions move out between ~1.5 and 5 solar radii, C and O first and heavier ions and Fe later, with details depending on the model. The slow (~400 km s$^{-1}$) solar wind, which most closely matches the abundances of SEPs, is the most difficult to model.

Shock waves accelerate higher-energy SEPs primarily in the region of ~2–3 solar radii (*e.g.* Reames 2009), sampling the ions at each radius, without waiting for each element's charge states to freeze in. While not entirely isothermal, the SEPs provide orthogonal information on the state of the region of solar-wind formation. In addition, several SEP events show the presence of two thermal components, as we shall see.

## 2. Individual SEP Events

The STEREO spacecraft were launched in 2006, just as solar cycle 23 was ending, and they were widely separated from Earth when the larger SEP events of cycle 24 became available (see https://stereo-ssc.nascom.nasa.gov/where.shtml for locations). We consider well-resolved individual events with sufficient SEP intensities for abundance measurements on two or more spacecraft. Such events are somewhat rare.

### *2.1 The Event of 19 January 2012*

Properties of the gradual SEP event of 19 January 2012 are shown in Figure 2. The panels show the time history of He, O, and Fe at the two spacecraft (lower left), of the enhancement in Fe/O (center left), and of the derived daily-averaged best-fit source plasma temperatures at *Wind* and STEREO B (upper left). Daily averages are required, especially for the much-smaller-geometry STEREO LET, to accumulate a sufficient sample of the rarer heavy ions. For each temperature measurement in the upper-left panel, a curve of $\chi^2/m$ *vs. T* that determined it is shown in the upper right panel with the corresponding symbol and color. The lower-right panel samples three days of best-fit enhancements *vs. A/Q* data and fits for STEREO B. The locations of the two spacecraft relative to the CME longitude are shown above the figure with spiral field lines shown for the typical solar wind speed early in the event.

The detailed analysis procedure used for Figure 2 and related figures is as follows: For each spacecraft on each day the average abundances of the 10 or 11 elements are measured and divided by the average coronal abundances (Reames 2017; see also Reames 1995, 2014) to produce observed *enhancements*. Since *A/Q* varies with temperature, each set of observed enhancements is fitted against the corresponding set of log *A/Q* for *each* of the 11 logarithmically-spaced temperature values from 0.8 to 8 MK to obtain the slope and intercept in log-log space and the $\chi^2/m$ value of each fit. For the 20 January 2012 event in Figure 2 we analyze 2 spacecraft × 3 days × 11 temperatures = 66 fits. For each day and each spacecraft, a plot of $\chi^2/m$ *vs. T* is shown in the upper right panel of Figure 2. The best-fit temperature of the minimum (or minima) of $\chi^2/m$ *vs. T* for each spacecraft and day is plotted at the appropriate time in the upper left panel of Figure 2 using the same color (day) and symbol (spacecraft) that was used in the corresponding $\chi^2/m$ vs. $T$ plot (2 spacecraft × 3 days = 6 curves). Since it is not practical to plot all 66 fits of enhancement vs. $A/Q$, each involving 10 or 11 data points, the lower right panel shows selected best-fit lines and data of enhancement vs. $A/Q$; for Figure 2 this panel compares the best fits for STEREO B on each of 3 days, with colors corresponding to those in the upper-left and upper-right panels – showing minimal time dependence. The dates are also listed in the appropriate color in the lower-right panel. Analysis of other SEP events is displayed in figures similar to Figure 2.

Best-fit temperatures in Figure 2 are in the range of 1.3–1.6 MK at both spacecraft throughout the event. An exception is on 21 January at STEREO B where the plot of $\chi^2/m$ has double minima indicating two temperatures: one minimum at 1.6±0.4 MK and another at 4±1 MK. This may indicate the presence of two thermal components; we will see a better example in a later event. The LET at STEREO B has poor statistical measurements of Ar and Ca at this time, so the difference between 1.6 and 4 MK depends upon whether the elements N and O are Ne-like or He-like in their enhancements.

Apart from the 4-MK-measurement, the best-fit temperature at STEREO B is 1.6±0.2 MK while that at *Wind* is 1.26±0.15 MK, but this difference is just the spacing of the temperature grid and is hardly significant.

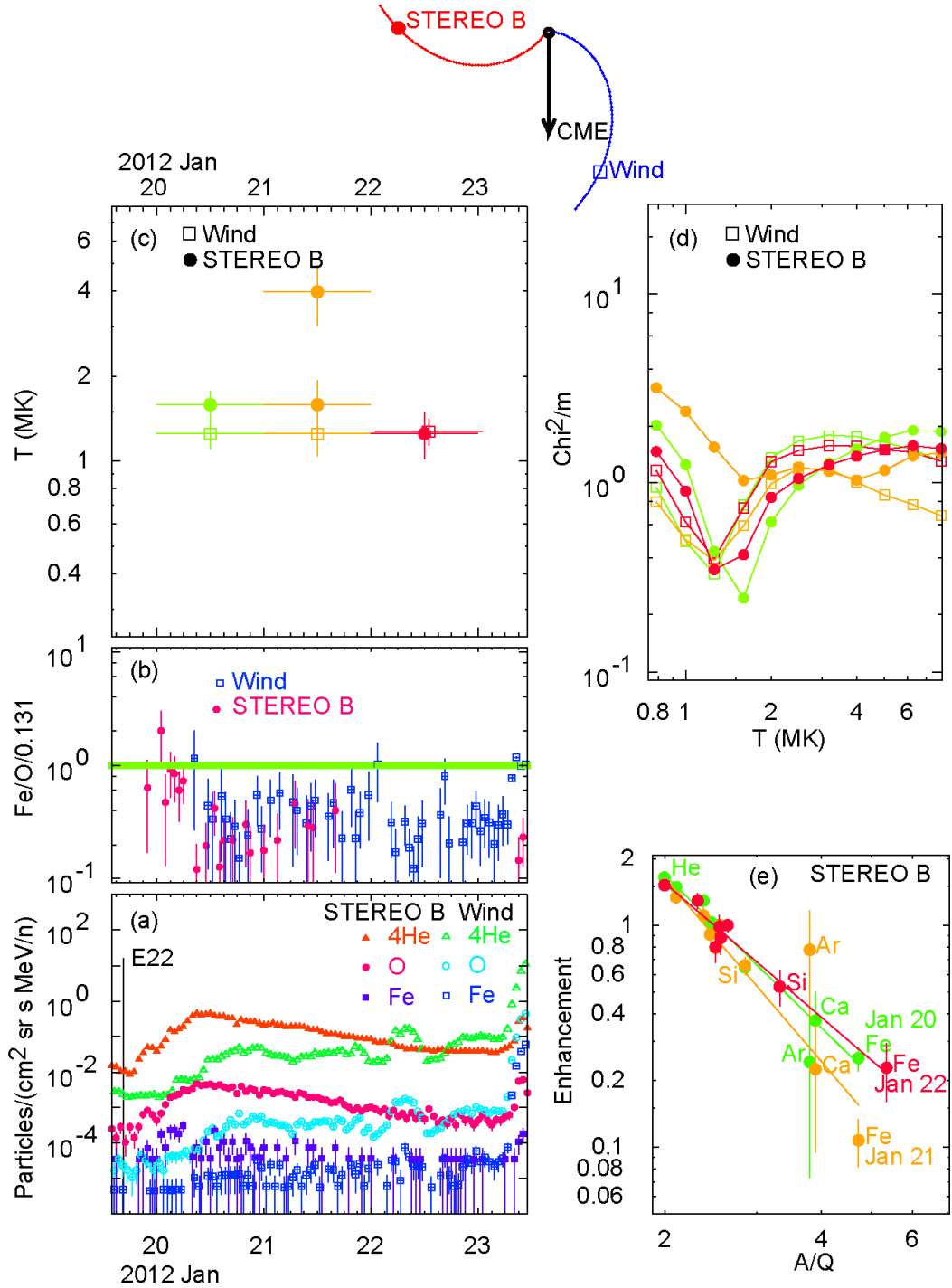

**Figure 2.** The 19 January 2012 SEP event is shown for STEREO-B (filled) and *Wind* (open symbols): (a) the intensities of He, O, and Fe *vs.* time, (b) the enhancements of Fe/O *vs.* time, (c) the derived source temperatures, *T vs.* time, (d) $\chi^2/m$ *vs. T*, color coded for each spacecraft and time, and, (e) enhancement *vs. A/Q* fits for three days at STEREO-B. The spacecraft locations relative to the CME are shown above.

While the quality of the fit of enhancement *vs. A/Q* can be judged from $\chi^2$ and from the sample fits in the lower-right panel of Figure 2, it is instructive to show how the measured enhancements map onto the theoretical distribution of *A/Q vs. T* at the best-fit temperature. Such a plot is shown as Figure 3. The mapping uses the best-fit slope and intercept for that spacecraft on that day to relate log (X/O) to log (*A/Q*). Similar comparisons were shown during a large number of events for data from *Wind* (Reames 2016a).

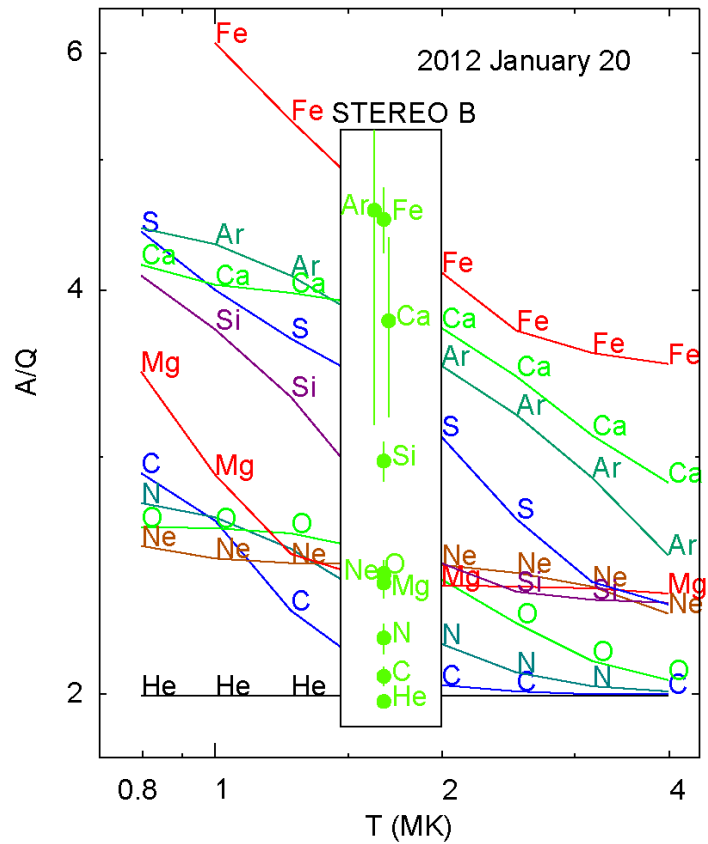

**Figure 3**. Measured values of enhancements, X/O at STEREO-B on 20 January 2012, are mapped into the theoretical *A/Q vs. T* at 1.6 MK using the best-fit parameters.

## *2.2 The Event of 23 January 2012*

The event of 23 January 2012 immediately follows the one that we have just considered, and it is large enough to provide adequate measurements on all three spacecraft. Figure 4 shows a comparison and analysis of *Wind* and STEREO-A while Figure 5 compares *Wind* and STEREO-B and shows an analysis of STEREO-B data.

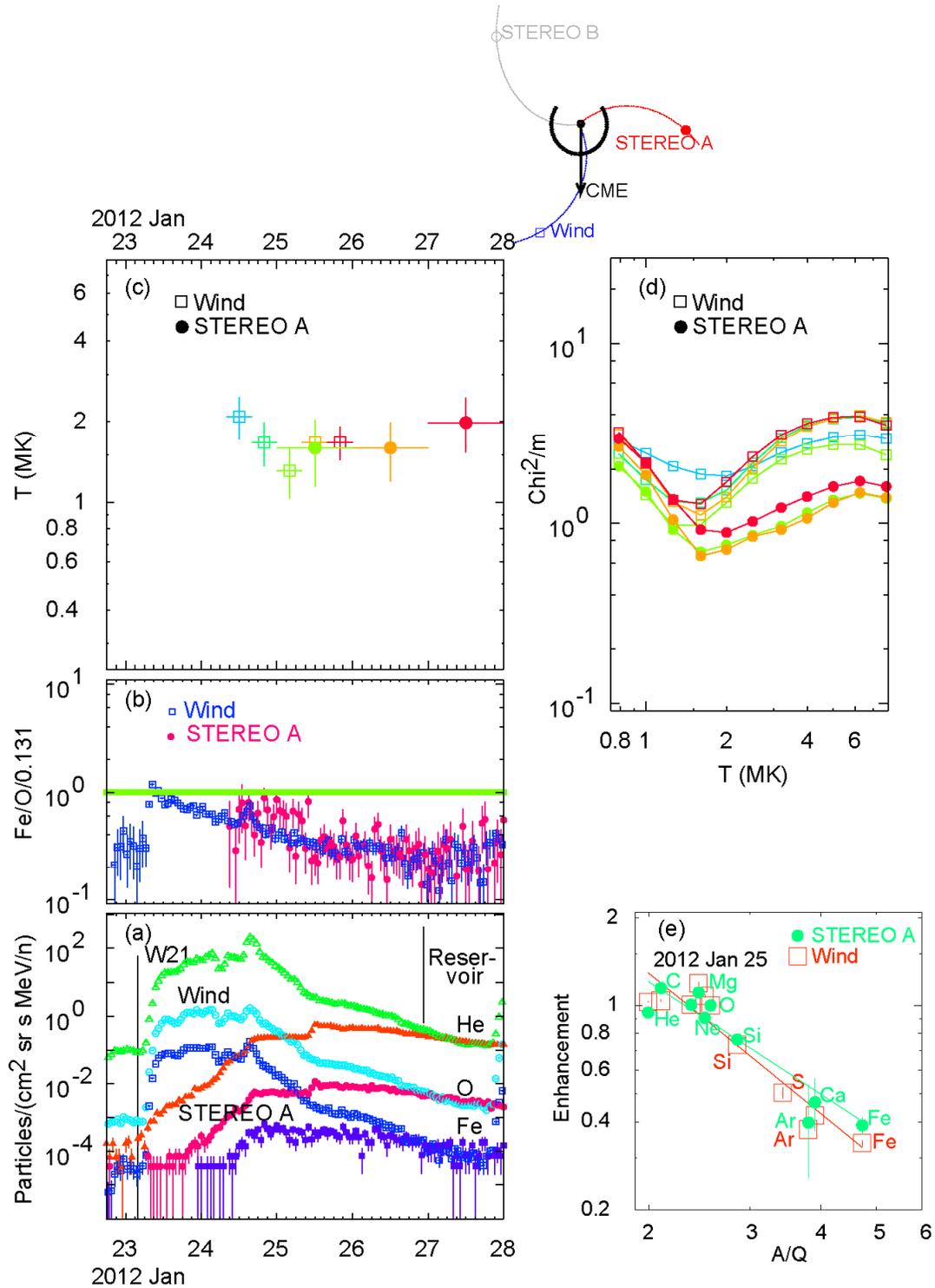

**Figure 4.** The 23 January 2012 SEP event is shown for STEREO-A (filled) and *Wind* (open symbols): (a) the intensities of He, O, and Fe *vs.* time, (b) the enhancements of Fe/O *vs.* time, (c) the derived source temperatures, $T$ *vs.* time, (d) $\chi^2/m$ *vs.* $T$, color coded for each spacecraft and time, and, (e) enhancement *vs.* $A/Q$ fits are compared for STEREO-A and *Wind* on 25 January 2012. The spacecraft locations relative to the CME are shown above.

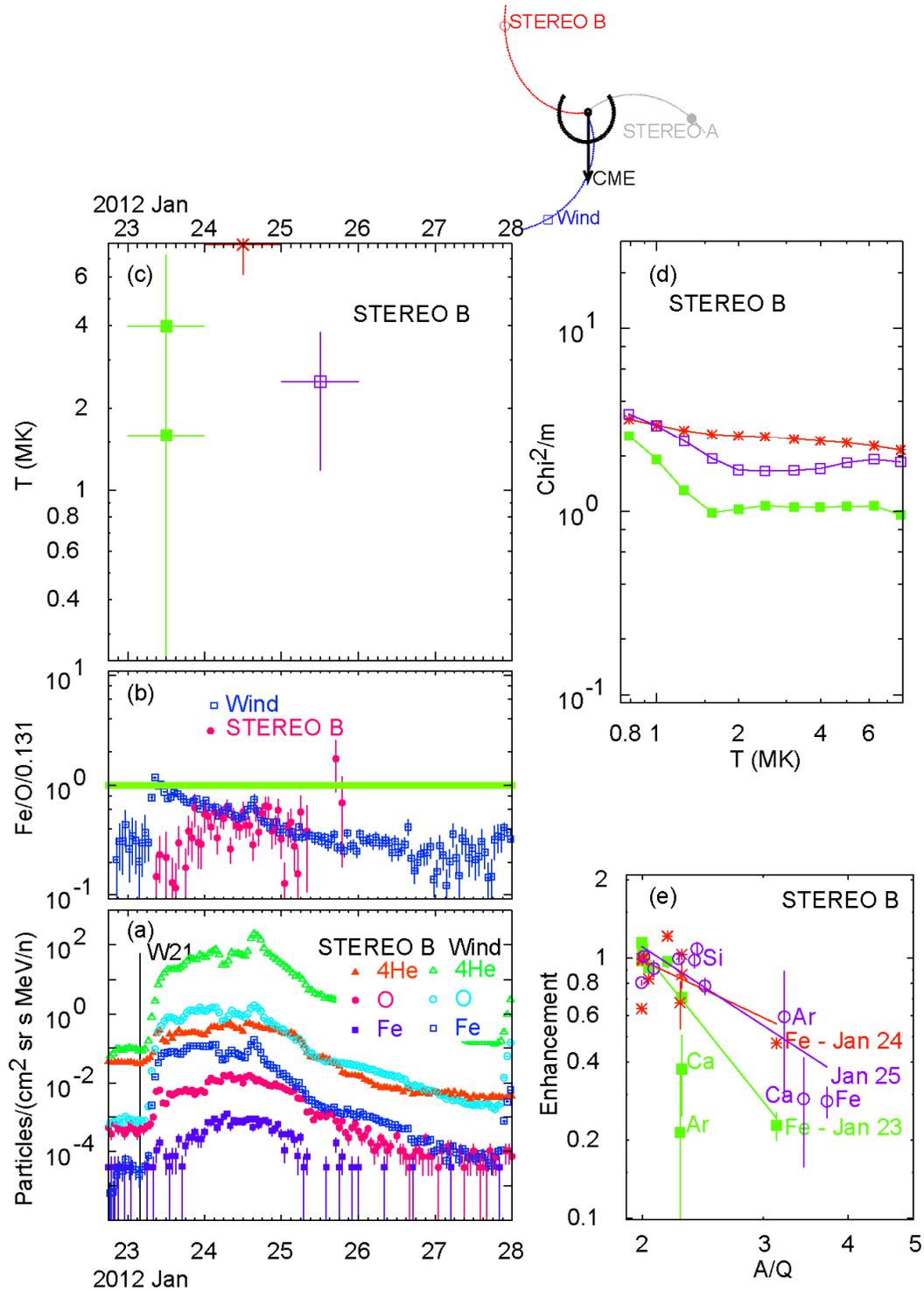

**Figure 5**. The 23 January 2012 SEP event is shown for STEREO-B (filled) and *Wind* (open symbols): (a) the intensities of He, O, and Fe *vs.* time, (b) the enhancements of Fe/O *vs.* time, (c) the derived source temperatures, *T vs.* time at STEREO-B, (d) $\chi^2/m$ *vs. T*, color coded for time at STEREO-B, and, (e) enhancement *vs. A/Q* fits for three days at STEREO-B. The spacecraft locations relative to the CME are shown above. Temperatures are poorly defined at STEREO-B for this SEP event.

The analysis of the data in Figure 4 shows a relatively consistent temperature ≈ 1.6±0.4 MK determined at both *Wind* and STEREO-A. The elements He and Mg do not fall on the fitted curve well, but the results are generally consistent.

When we get to STEREO-B in Figure 5, however, the comparison fails. The STEREO-B temperatures are erratic, the plots of $\chi^2$ are unconvincing, and the power-law behavior seems to be missing. STEREO-B is far to the east of *Wind* but the event timing clearly indicates we are observing the same event. What can be wrong?

In Figure 6 we show the directly observed enhancement *vs.* Z on each day at each spacecraft. These distributions all seem to peak at Mg and cannot really be power-law in *A/Q*. Enhancements are flat or rising with Z below Mg and falling above. However, all of these non-power-law distributions are similar in character and can be credibly associated with the same processes seen by each spacecraft. Note also that these distributions differ from the pre-event distributions which are those from the 19 January 2012 event.

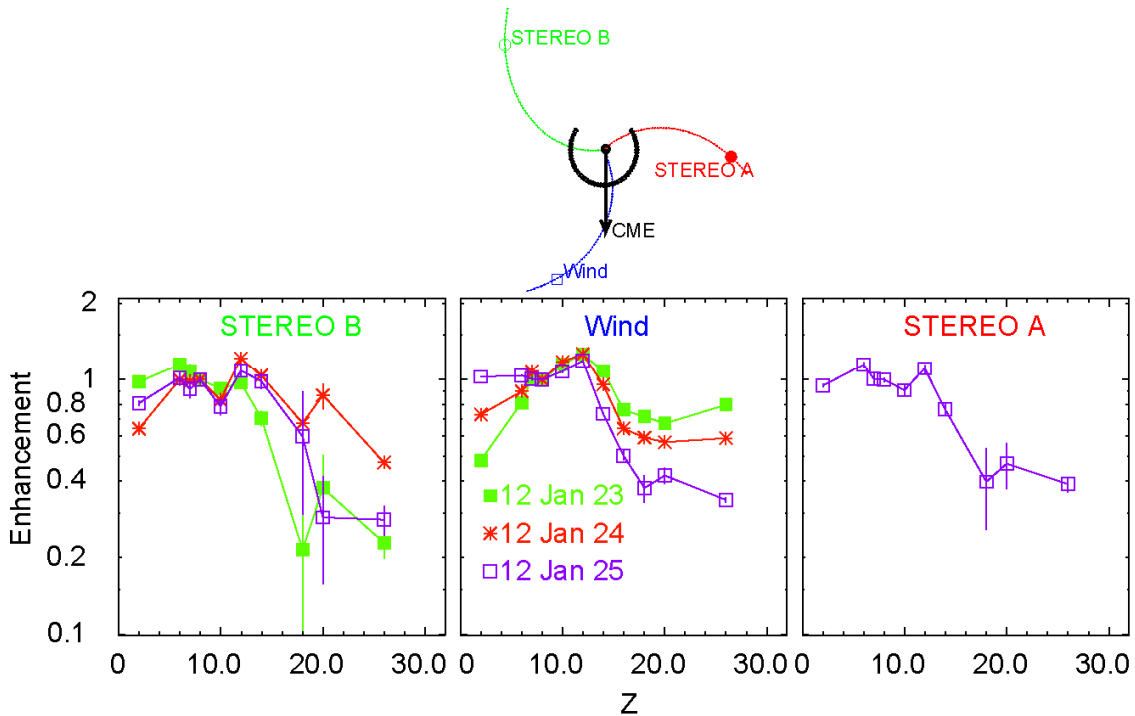

**Figure 6.** Enhancements of each element *vs.* Z on three days at each spacecraft. All of the distributions peak at Z = 12 (Mg) and cannot truly be power law in *A/Q*, yet they are all similar in this way.

Can the break at Mg be caused by some specific magnetic rigidity? To test this we compare the distributions of 2.5 – 3.2, 3.2 – 5 and 5 – 10 MeV amu$^{-1}$ at *Wind* in Figure 7.

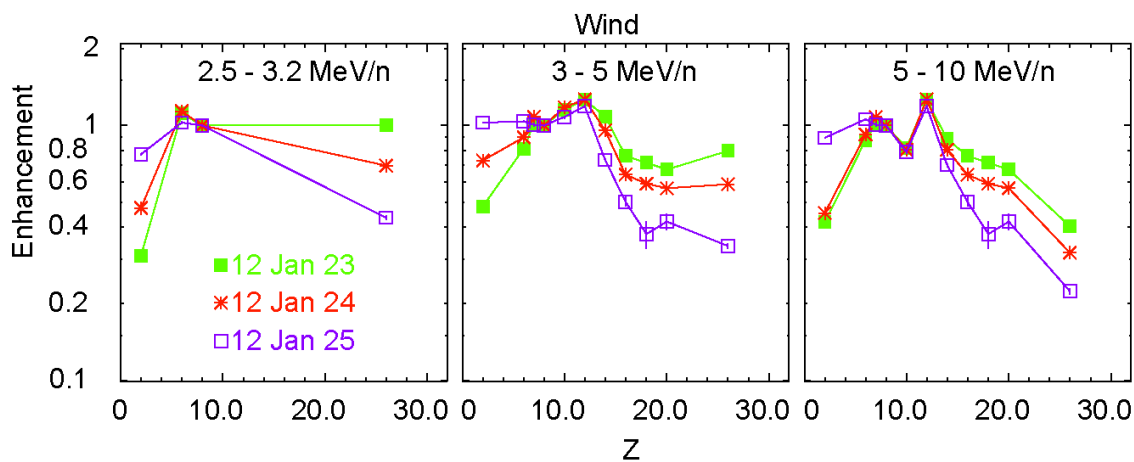

**Figure 7**. Enhancements *vs. Z* for three days are compared for three energy intervals at *Wind*.

The suppression of Fe increases with energy, and with time, in Figure 7. This may result from a spectral break at a break energy that varies as *Q/A*, as suggested by Tylka and Lee (2006). The break energy could begin at ~3 MeV amu$^{-1}$ for Fe, but above 10 MeV amu$^{-1}$ for elements Mg and below. In any case, this break destroys the temperature fit. Nevertheless, the entire 222° sweep of the event seems to have a similar unique enhancement-distribution pattern, one that was not present in the event that immediately preceded it.

### *2.3 The Event of 17 May 2012*

The SEP event of 17 May 2012 was seen at *Wind* and STEREO A and is shown in Figure 8. This event was a ground-level event (GLE) at Earth, meaning that the intensities of GeV protons exceed those of galactic cosmic rays and secondary particles from nuclear reactions in the atmosphere could be measured at ground level for this event. Coronagraph images of the CME and shock have been studied by Gopalswamy *et al.* (2013).

On 18 and 19 May 2012, *Wind* records temperatures of 1.0±0.5 MK while STEREO-A finds 1.6±0.2 MK, different, although not statistically significant. The suppression of Fe is fairly small at *Wind* so there is not much leverage *vs. A/Q* to allow measurement of *T*, and thus the minimum in $\chi^2$ is shallow. The suppression of Fe at STEREO-A is quite strong, but the intensities are not large.

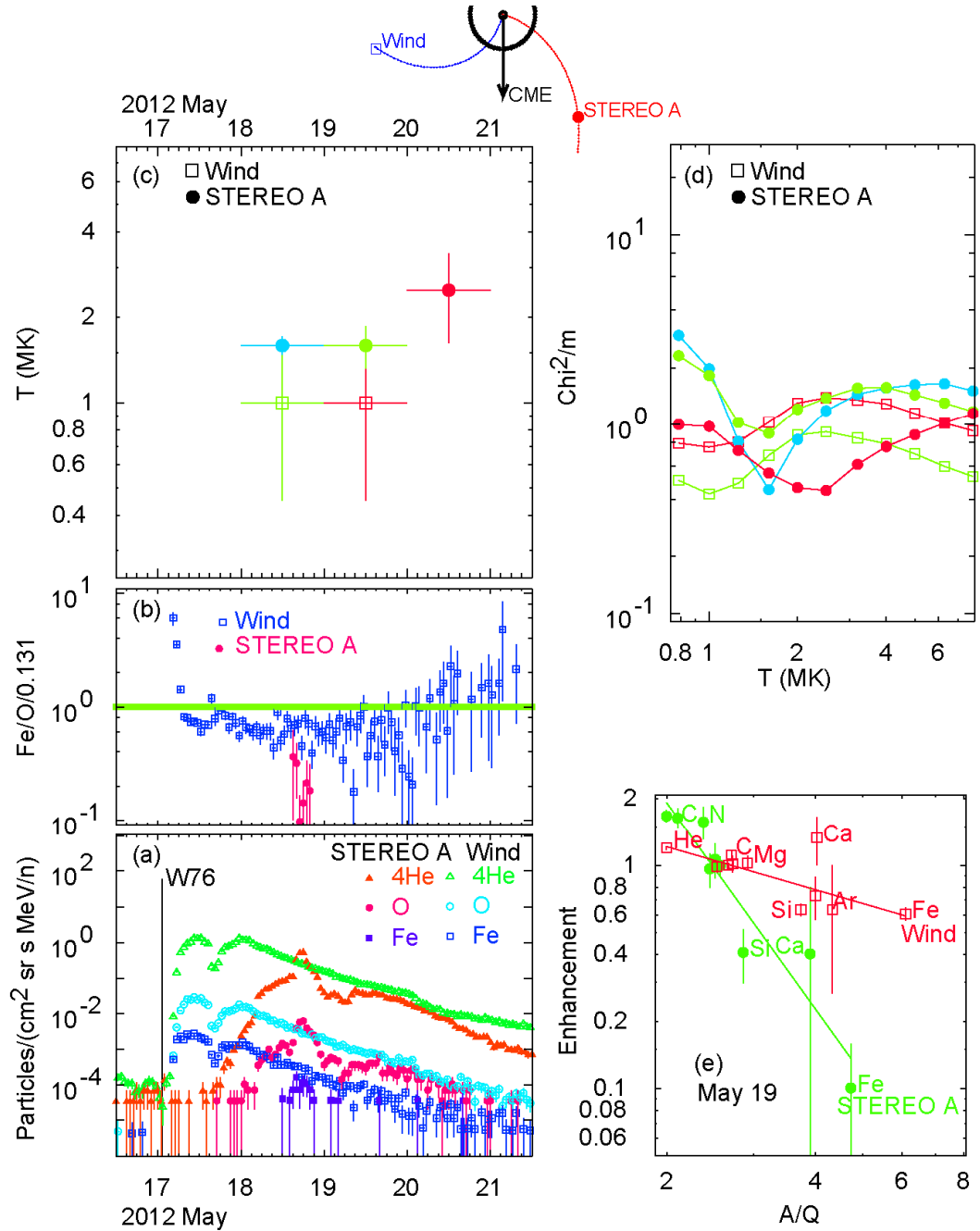

**Figure 8.** The 17 May 2012 SEP event is shown for STEREO-A (filled) and *Wind* (open symbols): (a) the intensities of He, O, and Fe *vs.* time, (b) the enhancements of Fe/O *vs.* time, (c) the derived source temperatures, $T$ *vs.* time, (d) $\chi^2/m$ *vs.* $T$, color coded for each spacecraft and time, and, (e) enhancement *vs.* $A/Q$ fits are compared for STEREO-A and *Wind* on 19 May 2012. The spacecraft locations relative to the CME are shown above.

### *2.3 The Event of 31 August 2012*

An analysis of the event of 31 August 2012 is shown in Figure 9.

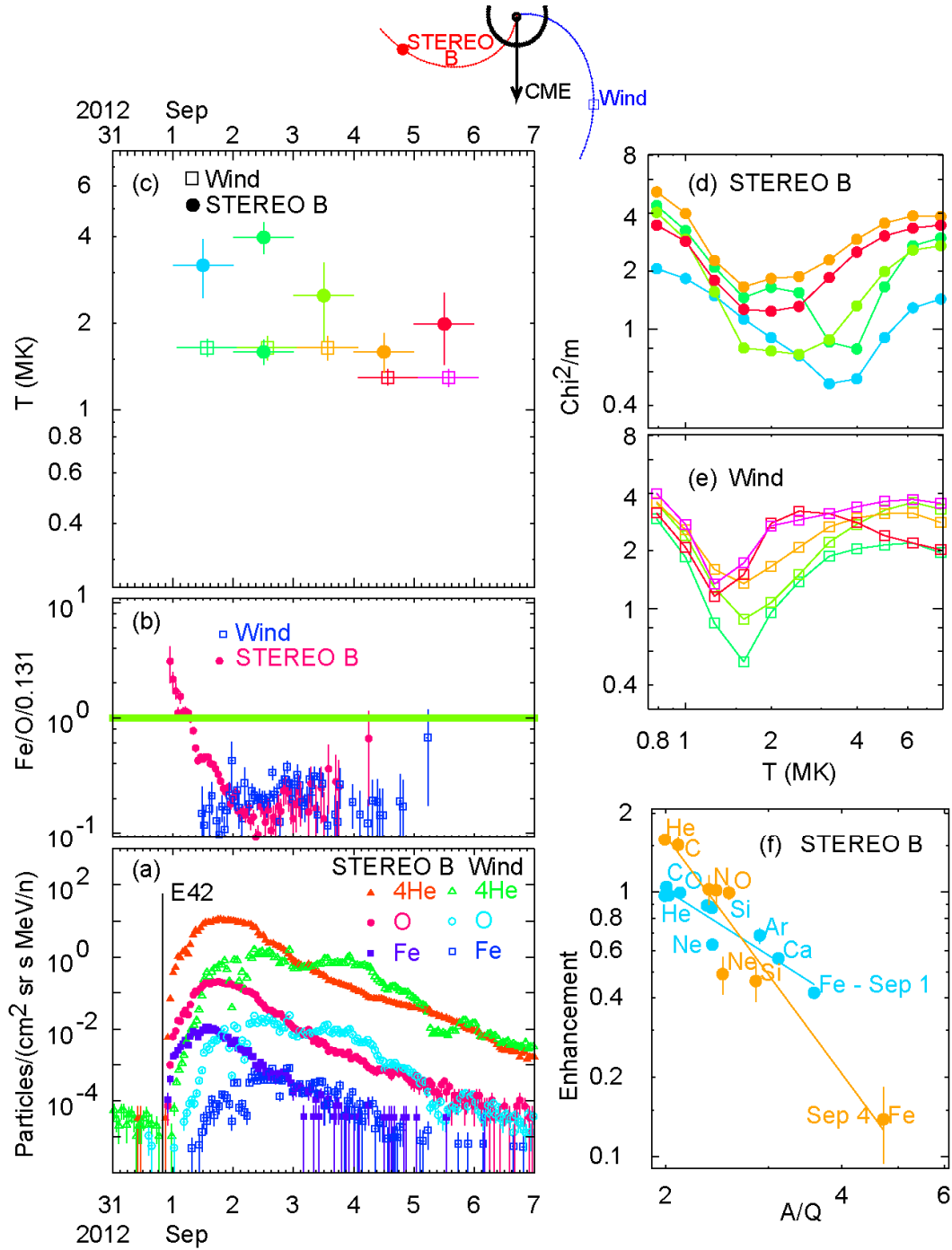

**Figure 9.** The 31 August 2012 SEP event is shown for STEREO-B (filled) and *Wind* (open symbols): (a) the intensities of He, O, and Fe *vs.* time, (b) the enhancements of Fe/O *vs.* time, (c) the derived source temperatures, *T vs.* time, (d and e) $\chi^2/m$ *vs. T*, color coded for time, for (d) STEREO-B and (e) *Wind*, and, (f) enhancement *vs. A/Q* fits for early and late days at STEREO-B. The spacecraft locations relative to the CME are shown above.

This event finally shows significant differences in temperature between the two spacecraft early in the event which fades away later. On the first three days of September 2012, temperatures found at *Wind* were consistently 1.6±0.2 MK followed by 1.26±0.1 MK on 4 September 2012. Meanwhile the temperature seen at STEREO-B begins at 3.2±0.8 MK on 1 September 2012, *shows double minima*, a strong one at 4.0±0.5 and a very weak one at 1.6±0.2 MK, on 2 September 2012, becomes 2.5±0.8 MK on 3 September, and finally comes to 1.6±0.3 MK on 4 September 2012, as seen in the upper-left panel of Figure 9.

Thus we see hot material, possibly impulsive suprathermals, at the eastern-most spacecraft and cooler material west of central meridian. As the CME emerges and the Sun rotates both particle populations are seen at STEREO-B, and then mostly cooler material. Both spacecraft may even be in a reservoir beginning near 1200UT on 2 September 2012, but the intensities are not exactly equal *and the temperatures differ*.

Note in the lower-right panel of Figure 9 that the patterns of enhancement at STEREO-B are much different on the 1$^{st}$ and 4$^{th}$ of September 2012. Not only the slopes of the fit, but O is near He on the 1$^{st}$ but well separated on the 4$^{th}$, corresponding to the different temperature. In Figure 10 we compare the patterns of enhancements at the two spacecraft on the 1$^{st}$ of September 2012 with the theoretical values.

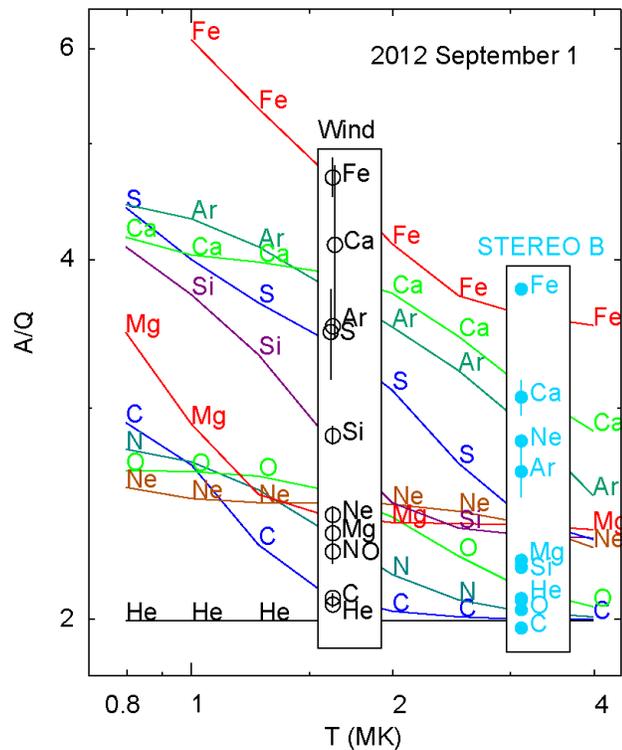

**Figure 10.** Measured values of enhancements, X/O at *Wind* and STEREO-B on 1 September 2012, are mapped into the theoretical plot *A/Q vs. T* at 1.6 MK and 3.2 MK, respectively, using the best-fit parameters. These measurements correspond to different source plasma temperatures at different locations in space at the same time during the SEP event.

Enhancements in the 1.6 MK plasma at *Wind* show: C down with He, N and O enhanced like Ne and Mg, but Si above them. At STEREO-B, C and O are actually below He, while Mg and Si are under-enhanced, yet Ne is over-enhanced. Large Ne enhancements, exceeding those of Mg and Si, are a signature of impulsive SEP events and impulsive suprathermal ions (Reames, Cliver, and Kahler 2014a; Reames 2015) although this enhancement of Ne is quite large.

## 3. Summary

In this study, we have found several well-resolved SEP events that were sufficiently extensive and intense that we could measure abundances of multiple elements, so as to define source plasma temperatures for several days at two or more spacecraft. We find the following:

i) One case, 31 August 2012, shows clear and persistent differences in abundances and source temperatures, 3.2±0.8 to the East and 1.6±0.2 to the West. The difference diminishes in three days, perhaps aided by CME expansion.

ii) In one case, 23 January 2012, the source temperature appears to be ~1.6 MK, but a strong non-power-law abundance component disrupts the measurement, especially at STEREO-B. Nevertheless, this unique abundance pattern is seen at all three spacecraft over 222° of solar longitude for three days.

iii) In the remaining two cases, abundances and temperatures are statistically consistent across a broad spatial distribution within ~25% measurement resolution. Another substantial spatial difference is possible, but cannot be proven. The event of 17 May 2012 shows a temperature that is 1.6±0.2 MK at STEREO A, to the west of the CME source but 1.0±0.5 MK at *Wind* to the east, but the errors overlap.

In general, we can only derive temperatures when there are abundance enhancements or suppressions, the larger the better, and not when abundances approach the reference abundances. We have also seen an example of non-power behavior, which disrupts the measurement.

The temperature that we determine is the electron temperature at the location wherever the dominant *A/Q*-dependence is determined. For temperatures <2 MK, it is the

Spatial Distribution of Abundances in SEP Events – Reames – accepted by Solar Physics

average electron temperature of the coronal plasma accelerated directly by the shock wave. For $2 < T < 4$ MK, it is likely to be the electron temperature of the plasma in the reconnection region of the jets that produced the impulsive suprathermal ions that were later accelerated by the shock. In either case, this technique begins to provide new information on SEP ionic charge states and temperatures that would be completely unavailable otherwise.

**Acknowledgments**: The author thanks Steve Kahler, Chee Ng, and Lun Tan for many helpful discussions about this manuscript.

## Disclosure of Potential Conflicts of Interest

The author declares he has no conflicts of interest.

## References

Breneman, H.H., Stone, E.C.: 1985, *Astrophys. J. Lett.* **299**, L57.
Cliver, E.W., Ling, A.G.: 2007, *Astrophys. J.* **658**, 1349.
Cliver, E.W., Ling, A.G.: 2009, *Astrophys. J.* **690**, 598.
Cliver, E.W., Kahler, S.W., Reames, D.V.: 2004, *Astrophys. J.* **605**, 902.
Cohen, C.M.S., Mason, G. M., Mewaldt, R.A., von Rosenvinge, T.T: 2013, *AIP Conf. Proc.* **1539**, 151.
Desai, M.I., Giacalone, J.: 2016, *Livi. Rev. of Solar Phys.*, doi: 10.1007/s41116-016-0002-5.
Desai, M.I., Mason, G.M., Dwyer, J.R., Mazur, J.E., Gold, R.E., Krimigis, S.M., Smith, C.W., Skoug, R.M.: 2003, *Astrophys. J.* **588**, 1149.
Desai, M.I., Mason, G.M., Wiedenbeck, M.E., Cohen, C.M.S., Mazur, J.E., Dwyer, J.R., Gold, R.E., Krimigis, S.M., Hu, Q., Smith, C.W., Skoug, R.M.: 2004, *Astrophys. J.* **661**, 1156.
Desai, M.I., Mason, G.M., Gold, R.E., Krimigis, S.M., Cohen, C.M.S., Mewaldt, R.A., Mazur, J.E., Dwyer, J.R.: 2006, *Astrophys. J.* **649**, 470.
Daibog, E. I., Stolpovskii, V. G., Kahler, S. W.: 2003 *Cosmic Res.* **41**, 128.
Gosling, J.T.: 1993, *J. Geophys. Res.* **98**, 18937.
Gopalswamy, N., Yashiro, S., Krucker, S., Stenborg, G., Howard, R.A.: 2004, *J. Geophys. Res.,* **109**, A12105.
Gopalswamy, N., Xie, H., Yashiro, S., Akiyama, S., Mäleka, P., Usoskin, I.G.: 2012 *Space Sci. Rev.* **171**, 23, DOI 10.1007/s11214-012-9890-4
Gopalswamy, N., Xie, H., Akiyama, S., Yashiro, S., Usoskin, I.G., Davila, J.M.: 2013, *Astrophys. J. Lett.* **765** L30.
Kahler, S.W.: 1992, *Annu. Rev. Astron. Astrophys*. **30**, 113.
Kahler, S.W.: 1994, *Astrophys. J*. **428**, 837.
Kahler, S.W.: 2001, *J. Geophys. Res*. **106**, 20947.

Kahler, S.W.: 2007 *Space Sci. Rev*. **129**, 359.

Kahler, S.W., Cliver, E.W., Tylka, A.J., Dietrich, W.F.: 2012 *Space Sci. Rev*. **171**, 121.

Kahler, S.W., Sheeley, N.R.Jr., Howard, R.A., Koomen, M.J., Michels, D.J., McGuire R.E., von Rosenvinge, T.T., Reames, D.V.: 1984, *J. Geophys. Res*. **89**, 9683.

Klecker, B.: 2013; *J. Phys.: CS*-**409**, 012015.

Lee, M.A.: 1997, In: Crooker, N., Jocelyn, J.A., Feynman, J. (eds.) *Coronal Mass Ejections, Geophys. Monograph* **99**, AGU, Washington, 227.

Lee, M.A.: 2005, *Astrophys. J. Suppl.,* **158**, 38.

Lee, M.A., Mewaldt, R.A., Giacalone, J.: 2012, *Space Sci. Rev.,* **173**, 247.

Luhmann, J. G., Curtis, D.W., Schroeder, P., McCauley, J., Lin, R.P., Larson, D.E.. *et al*.: 2008 *Space Sci. Revs*. **136**. 117. DOI: 10.1007/s11214-007-9170-x

Mazzotta, P., Mazzitelli, G., Colafrancesco, S., Vittorio, N.: 1998 *Astron. Astrophys Suppl*. **133**, 403.

Mason, G.M.: 2007 *Space Sci. Rev*. **130**, 231.

Mason, G.M., Gloeckler, G., Hovestadt, D.: 1984, *Astrophys. J*. **280**, 902.

Mason, G.M., Mazur, J.E., Dwyer, J.R.: 1999, *Astrophys. J. Lett*. **525**, L133.

McKibben, R. B.: 1972, *J. Geophys. Res*. **77**, 3957.

Mewaldt, R.A., Cohen, C.M.S., Cook, W.R., Cummings, A.C., Davis, A.J., Geier, S., Kecman, B., Klemic, J., Labrador, A.W. Leske, R.A., *et al.*: 2008, *Space Sci. Rev*. **136**, 285.

Meyer, J. P.: 1985, *Astrophys. J. Suppl*. **57**, 151.

Ng, C.K., Reames, D.V., Tylka, A.J.: 2003, *Astrophys. J*. **591**, 461.

Post, D.E., Jensen, R.V., Tarter, C.B., Grasberger, W.H., Lokke, W.A.: 1977, *At. Data Nucl. Data Tables*, **20**, 397

Reames, D.V.: 1995, *Adv. Space Res.* **15** (7), 41.

Reames, D.V.: 1999, *Space Sci. Rev.,* **90**, 413.

Reames, D.V.: 2009, *Astrophys. J*. **706**, 844.

Reames, D.V.: 2013, *Space Sci. Rev*. **175**, 53.

Reames, D.V.:2014, *Solar Phys*. **289**, 977, DOI: 10.1007/s11207-013-0350-4

Reames, D.V.: 2015, *Space Sci. Rev.,* **194**: 303, DOI: 10.1007/s11214-015-0210-7. (arXiv: 1510.03449).

Reames, D.V.: 2016a, *Solar Phys*., **291** 911, DOI: 10.1007/s11207-016-0854-9, (arXiv: 1509.08948).

Reames, D.V.:2016b, *Solar Phys*, **291** 2099, DOI: 10.1007/s11207-016-0942-x, (arXiv: 1603.06233).

Reames D.V.: 2017, *Solar Energetic Particles*, Springer, ISBN 978-3-319-50870-2, DOI 10.1007/978-3-319-50871-9.

Reames, D. V., Ng, C. K.: 2010, *Astrophys. J*. **722**, 1286.

Reames, D.V., Barbier, L.M., Ng, C,K.:1996, *Astrophys. J*. **466**, 473.

Reames, D.V., Cliver, E.W., Kahler, S.W.: 2014a, *Solar Phys*. **289**, 3817, DOI: 10.1007/s11207-014-0547-1 (arXiv: 1404.3322)

Reames, D.V., Cliver, E.W., Kahler, S.W.: 2014b, *Solar Phys*. **289**, 4675, DOI: 10.1007/s11207-014-0589-4 (arXiv:1407.7838)

Reames, D.V., Cliver, E.W., Kahler, S.W.: 2015, *Solar Phys*. **290**, 1761, DOI: 10.1007/s11207-015-0711-2 (arXiv: 1505.02741).

Reames, D.V., Kahler, S.W., Ng, C.K.:1997, *Astrophys. J*. 491, 414.

Reames, D.V., Meyer, J.P., von Rosenvinge, T.T.: 1994, *Astrophys. J. Suppl*., **90**, 649.

Reames, D.V., Ng, C.K., Tylka, A.J.: 2012, *Solar Physics* **285**, 233, DOI 10.1007/s11207-012-0038-1.

Roelof, E.C., Gold, R.E., Simnett, G.M., Tappin, S.J., Armstrong, T.P., Lanzerotti, L.J.: 1992  *Geophys. Res. Lett.* **19**, 1247.

Rouillard, A.C., Odstrčil, D., Sheeley, N.R. Jr., Tylka, A.J., Vourlidas, A., Mason, G., Wu, C.-C., Savani, N.P., Wood, B.E., Ng, C.K., *et al*.: 2011, *Astrophys. J.* **735**, 7.

Rouillard, A., Sheeley, N.R.Jr., Tylka, A., Vourlidas, A., Ng, C.K., Rakowski, C., Cohen, C.M.S., Mewaldt, R.A., Mason, G.M., Reames, D., *et al.:* 2012, *Astrophys. J.* **752**:44.

Tylka, A.J., Lee, M.A.: 2006, *Astrophys. J.* **646**, 1319.

Tylka, A.J., Cohen, C.M.S., Dietrich, W.F., Lee, M.A., Maclennan, C.G., Mewaldt, R.A., Ng, C.K., Reames, D.V.: 2005, *Astrophys. J.* **625**, 474.

von Rosenvinge, T.T., Barbier, L.M., Karsch, J., Liberman, R., Madden, M.P., Nolan, T., Reames, D.V., Ryan, L., Singh, S. *et al*.: 1995, *Space Sci. Rev.* **71**, 155.

Wang, L., Lin, R.P., Krucker, S., Mason, G.M.: 2012 *Astrophys. J.* **759**, 69.

Webber, W. R.: 1975, *Proc. 14$^{th}$  Int. Cos. Ray Conf*, (Munich), **5**, 1597